\documentclass[aps,12pt,final,notitlepage,nobibnotes,nofootinbib,
superscriptaddress,noshowpacs,
centertags,epsfig]{revtex4}

\newcommand{\be}{\begin{equation}}
\newcommand{\ee}{\end{equation}}
\newcommand{\<}{\langle}
\renewcommand{\>}{\rangle}
\newcommand{\reff}[1]{(\ref{#1})}

\begin{document}

\title{Path integrals and degrees of freedom in many-body systems and relativistic
field theories}


\author{
  { Fabrizio Palumbo~\thanks{This work has been partially 
  supported by EEC under the contract HPRN-CT-2000-00131}}             \\[-0.2cm]
  {\small\it INFN -- Laboratori Nazionali di Frascati}  \\[-0.2cm]
  {\small\it P.~O.~Box 13, I-00044 Frascati, ITALIA}          \\[-0.2cm]
  {\small e-mail: {\tt palumbof@lnf.infn.it}}     
   }


\thispagestyle{empty}   

\begin{abstract}

The identification of physical degrees of freedom is sometimes obscured in
the path integral formalism, and this makes it difficult to impose some constraints or to do
some approximations.
 I review a number of cases where the difficulty is overcame by deriving the path integral from the
operator form of the partition function after such identification has been made.



\end{abstract}

\maketitle



\section{Introduction}

It is a great pleasure for me to contribute to this volume in honor of Prof. Yu. A. Simonov.
This is an occasion for me to remember again the origin of our friendship which extended over the 
years well beyond what it would appear from our joint papers. 

The subject of my contribution is focussed on my recent interests in several problems which have a common
feature: The identification of degrees of freedom in a path integral. Indeed there are many situations
where such identification is helpful or necessary.  An old example is the thermodynamics of gauge theories 
~\cite{Dola}, but recently I met with many others. The first I will discuss here is how to find 
actions exactly equivalent to the standard ones but closer to the continuum at finite lattice
spacing\cite{Palu}. This includes the definition of the couplings of the chemical potential, an issue
of particular importance in QCD~\cite{Palu1}. The chemical potential is used to fix the expectation value 
of some charge operator. Fixing instead the value of the charge, namely selecting a specific charge sector
is somewhat more difficult, but interesting pieces of information can be obtained by an expansion 
in a given charge sector of the fermion determinant in series of the number of fermions\cite{Palu2}. The
last problem I will consider is how to describe the low lying excitations of fermionic systems, both
relativistic and nonrelativistic, by means of effective bosons, in short how to bosonize them~\cite{Palu3}.
All these problems require the  identification of the relevant degrees of freedom. In the last case to
determine the  structure  of the composite bosons in terms of the fermionic constituents, in the first
case to show the equivalence of different actions.

Now the path integral formalism is widely used because of its flexibility and the possibility of numerical
applications, but the identification of degrees of freedom is not always easy, while it can be 
conveniently achieved in the operator form of the partition function. Therefore in all the above
problems I will first identify the relevant degrees of freedom in the Fock space where the partition
function is defined, then I will introduce the appropriate constraints or approximations and finally I
will derive the constrained or approximated path integral.

As I said all the subjects I mentioned have a common feature in the role played by the 
identification of degrees of freedom, but are otherwise very different. Therefore the motivations for
their investigation are given separately in the relative Sections. In Section 2 I will report the results I
will use later about the standard derivation of the path integral from the operator form of the partition
function. In Section 3 I will show how to derive an action different from the standard one and closer
to the continuum in the nonrelativistic case. In Section 4 I will carry out the corresponding
derivation for relativistic field theories, confining myself to the couplings of the chemical potential. In
Section 5 I will discuss the case of a given charge sector. In Section 6 I will present a general
method of  bosonization valid for relativistic and nonrelativistic theories and in Section 7 my
conclusions.

\section{The standard derivation of the Euclidean path integral from the partition function}

Let me introduce some definitions. I denote by  $\tau$ the temporal lattice spacing, by $N_0$ 
the number of temporal sites, by {\it x}$_0$ or $t$ the temporal component of the site position vector
$x$,  by $T$ the temperature, by $\mu$ the chemical potential, by $\hat{Q}$ the (electric, baryon...)
charge operator, and by $ \hat{{\cal T}}(x_0)$ the transfer matrix. 

In the nonrelativistic case ${\cal T}$ is expressed in terms of the Hamiltonian 
\be
{\cal T} = \exp \left( -\tau H \right),
\ee
and the Hamiltonian is the generator of continuous time translations. In relativistic theories instead
only the transfer matrix is known in general, and the above equation can be used to define a Hamiltonian,
but only as the generator of discrete translations by the time spacing $\tau$. Another important difference
is that the nonrelativistic interactions are generally quartic in the fields while the relativistic ones
are quadratic. Both features contribute to make the derivation of the euclidean path integral different
in the two cases.

$\hat{{\cal T}} $ is defined in terms of particle-antiparticle  creation-annihilation operators  
$\hat{c}^{\dagger}, \hat{d}^{\dagger},\hat{c},\hat{d} $ acting in a Fock space. It depends  on the 
time coordinate $x_0$ only through the dependence on it of other fields (for instance gauge fields). 
In fact the creation and annihilation  operators do not depend on $x_0$. They depend on the spatial 
coordinates ${\bf x}$ and on the internal quantum numbers (Dirac, color and flavor indices in the case 
of QCD), comprehensively represented by $I,J...$.

In the transfer matrix formalism often one has to do with quantities at a given (Euclidean) time  
$x_0$. For this reason we adopt a summation convention over spatial coordinates and intrinsic 
indices at fixed time. So for instance, for an arbitrary matrix $M$, we will write
\begin{equation}
 \hat{c}^{\dagger} M(x_0) \, \hat{c}
= \sum_{{\bf x},{\bf y},I,J} \hat{c}^{\dagger}_{{\bf x},I}
M_{{\bf x},I;{\bf y},J}(x_0) \hat{c}_{{\bf y},J}.
\end{equation}
In this notation the charge operator $ \hat{Q} $ has the form
\begin{equation} 
\hat{Q} = \hat{c}^{\dagger} \hat{c} - \hat{d}^{\dagger} \hat{d}. \label{charge}
\end{equation}

The grand canonical partition function can be written as a 
time-ordered product
\begin{equation}
Z = \mbox{Tr} \left\{  \exp \left(  {\mu \over T} \, \hat{Q} \right) 
\prod_{x_0}\hat{{\cal T}}(x_0) \right\},  \label{trace} 
 \end{equation}
which, using the relation $T^{-1}= \tau N_0$, and assuming the conservation of $\hat{Q} $ is conveniently
rewritten as
\begin{equation}
Z = \mbox{Tr} \left\{   \prod_{x_0} \left[\hat{{\cal T}}(x_0)
 \exp \left( \mu \, \tau \hat{Q} \right) \right]\right\}.  \label{trace1} 
 \end{equation}
           
The standard way~\cite{Nege,Lusc} to obtain the path integral form of $Z$
is to write all the operators in normal order and introduce between the factors in Eq.(\ref{trace1})
the identity
\be
{\cal I} = \int [dc^+ dc \,\, d d^+ d d] 
\exp(-c^+ c - d^+ d  ) |c d \> \< c d|,
\ee
where the basis vectors are coherent states
\begin{equation}
|c d \> = |\exp( - c \,\hat{c}^{\dagger} - d \,\hat{d}^{\dagger}) \>.
\end{equation}
 The $c^+,c, d^+,d $ are holomorphic/Grassmann variables and satisfy periodic/antiperiodic boundary
conditions in time for bosons/fermions respectively~\cite{Nege}. They have the label of the time slice 
where the identity operator is introduced. For the other indices they are subject to the same
convention as the creation and annihilation operators. The main property of coherent states is that they
are eigenstates of the annihilation operators 
\begin{equation}
\hat{c} |c d \> = c \, |c d \>.
\end{equation}
To get the functional form of the partition function it is only necessary to evaluate the matrix elements
$\<c_1 d_1| {\cal T}| c d\>$. This, as anticipated in the Introduction, must be done in different ways for
nonrelativistic and relativistic theories.

The first case appears more difficult because of the quartic interactions, but since in general there are
no ultraviolet divergencies (a detailed discussion of the departure from the standard form of the path
integral in the presence of singular potentials can be found in ~\cite{Klei}) we can make without any
error the approximation
\be
\exp \, (-\tau \hat{H}) \sim 1 - \tau \hat{H}.
\ee
 Let us then consider a many-body system with the Hamiltonian written in normal form
\be
\hat{H}(\hat{a}^{\dagger}, \hat{a}) = \sum_{{\bf x},{\bf y}} \left\{\hat{a}_{{\bf x}}^{\dagger} h_{{\bf x},
{\bf y}}
\hat{a}_{{\bf y}}  + 
\hat{a}_{{\bf x}}^{\dagger}  \hat{a}_{{\bf y}}^{\dagger}  v_{ {\bf x},{\bf y}} \hat{a}_{{\bf
y}}\hat{a}_{{\bf x}}  \right\}.
\ee
The standard expression of the euclidean path integral associated to this Hamiltonian is
\be
Z=\int [da^*da] \exp (-S)
\ee
where  
\be
S= \tau \sum_{t=1}^{N_0} \left\{ a^*_{t+1} \nabla_t a_t + H(a_{t+1}^*,a_t) \right\}
\ee
is the action and I denoted by
\be
\nabla_t f_t= { 1 \over \tau} (f_{t+1}-f_t)
\ee
the right discrete time derivative. Notice the time splitting between the fields and their 
conjugates, which implies a departure from the classical expression not only in derivative, but also
in potential terms.
Needless to say, neglecting this time splitting introduces finite errors~\cite{Nege}.

In the relativistic case the transfer matrix can be written\cite{Lusc}
\be
{\cal T}_{x_0} = T^{\dagger}_{x_0} T_{x_0}
\ee
where
\be
T = \exp \left( {\hat c}^{\dagger} M {\hat c} + {\hat d}^{\dagger} M^T {\hat d} \right)
  \,\exp\left( {\hat c} \, N \, {\hat d} \right)
\ee
where the upper script $T$ means transposed and I do not need to specify the matrices $M,N$. The 
matrix elements of ${\cal T}$ can be exactly evaluated~\cite{Lusc}, yielding the standard form of the
Euclidean path integral.

\section{Antinormal ordering: a different action in the nonrelativistic case} 

In the standard form of the action the time is split in the fields and their conjugates. This
is an artifact which makes the equations unnecessarily different from the continuous,
somewhat more complicated and somewhat confusing. In gauge theories, for instance, such
time splitting introduces a coupling of the chemical potential to the temporal gauge fields
which has been considered of physical significance, while it can be altogether avoided
in a different derivation of the path integral. I will first illustrate such a derivation 
in the nonrelativistic case~\cite{Palu1}. 

  Instead of the normal order I write the Hamiltonian in antinormal order 
(all the annihilation operators to the left of the creation ones)
\be
\hat{H} =  \sum_{{\bf x},{\bf y}} \left\{ h_0 +\hat{a}_{{\bf x}} h'_{{\bf x}, {\bf y}} \hat{a}_{{\bf
y}}^{\dagger}  + 
\hat{a}_{{\bf x}} \hat{a}_{{\bf y}} v_{ {\bf x},{\bf y}} \hat{a}_{{\bf
y}}^{\dagger}   \hat{a}_{{\bf x}}^{\dagger}   \right\}.
\ee
where
\begin{eqnarray}
h_0 &=& N^3 \left( - \sigma h_{{\bf x}, {\bf x}} + w \right)
\nonumber \\
h'_{{\bf x}, {\bf y}} & = &  \sigma h_{{\bf x}, {\bf y}} - w \delta_{{\bf x}, {\bf y}} . \label{rearr}
\end{eqnarray}
In the above equations
\be
w= \sigma v_{{\bf x}, {\bf x}} + \sum_{{\bf z}} v_{{\bf x}, {\bf z}} ,
\ee
 $N^3$ is the number of spatial sites and $ \sigma = -1$ for fermions and $+1$ for bosons. I assume that
the potential is sufficiently regular for $w$ to exist,  otherwise a regularization is needed or a more
drastic change in our procedure. Then I expand the transfer matrix and in each term I insert the identity
between the creation and annihilation operators. For the rightmost factor before taking the trace I move
the creation operators to the leftmost position. In this way the term with the time derivative remains
unchanged, but in the other terms all the fields appear at the same time 
\be
S' =  \tau \sum_t \left\{  h_0 + a^*_{t+1} \nabla_t a_t + (h'-h)  a^*_t a_t  \right.
 \left. + H(a^*_t,a_t) \right\}.
\ee
We can easily check on solvable models that this action gives the right results and that the
terms arising from the rearrangement in antinormal ordering cannot be neglected.

In the case of fermions the time splitting between the fields and their conjugates in the potential terms
can be avoided in the same way as for bosons, but also in a simpler way. In fact the Grassmann fields,
unlike the holomorphic variables, are independent from their conjugates, so that the simple transformation
\be
a_{t+1}^*  \rightarrow a^*_t
\ee
eliminates the time splitting everywhere, with the obvious exception of the term with the time
derivative, which is changed into the left one. 

Because in this case we have two different derivations of the path integral, by their comparison we can
get  nontrivial identities.

\section{Antinormal ordering: a different coupling of the chemical potential in
relativistic field theories}

In relativistic field theories the artificial time splitting affects only the coupling of the
chemical potential. In the Hasenfratz--Karsch--Kogut formulation\cite{Hase}, which is the standard one, 
for Wilson fermions such coupling takes the form
\be
\delta \, S = 2 K \sum_x \overline{q} \left\{ \left[ \exp \mu -1 \right] 
 P_0^{(+)} U_0 T_0^{(+)} \right.
 \left. \,\,\, + \left[ \exp(-\mu) -1 \right]  P_0^{(-)} T_0^{(-)}
  U_0^{(+)}  \right\} q
\ee
where $q$ is the quark field, K is the hopping parameter, $U_0$ the temporal link variable and
\begin{eqnarray}
P_0^{(\pm)}& =&{ 1\over 2} \left( 1\!\!1 \pm \gamma_0 \right)
\nonumber\\
T_0^{(\pm)}f(x_0) &=& f(x_0 \pm 1).
\end{eqnarray}
Because of gauge invariance, in the presence of the time splitting a coupling with the temporal
links is needed, and this led to the conclusion that a nonvanishing contribution of the chemical 
potential must necessarily involve a Polyakov loop. But Creutz showed in a toy model\cite{Creu} that this
is not true, and I will report~\cite{Palu2} in the sequel how to avoid such artifact in the full fledged
QCD. 

I write the exponential of the charge in the following way
\be
\exp (\mu \, a_0 \, \hat{Q})  = \int [dc^+ dc \,\, d d^+ dd] 
 \,\,\,\,  \exp (\delta S -c^+ c -d^+d)|cd \>\< cd|   \label{expQ}
\ee
where
\be
\delta S = \left( 1- \cosh( \mu a_0)\right)(c^+ c + d^+ d) 
 + \sinh( \mu a_0)(c^+ c - d^+ d).   \label{schem}
\ee
 The  expression of $\delta S $ is obtained by expanding the
exponential of the charge operator, putting all the terms in antinormal form, inserting in each monomial 
the unity between the set of annihilation and the set of creation operators and replacing them by their
Grassmannian eigenvalues. For the rightmost  exponential of the charge  before taking the trace one has
to  move the creation operators to the left of all the operators appearing under trace.

After this, the construction of the path integral proceeds in the standard way, and we get the standard
action with the exception of the coupling of the chemical
potential where all the fields appear at the same time
\be
 \delta S = \sum_x  \overline{q}   \big[ \left( 1- \cosh (\mu a_0)\right) 
+ \sinh (\mu a_0) \gamma_0 \big] B q . 
\ee
Here I only need to say that the matrix $B$ does not depend on $U_0$. What is important is to 
notice that the quark field and its conjugate are at the same time, and then the temporal 
Wilson variable disappeared.

\section{Expansion of the fermion determinant in the number of fermions in a given charge sector}

The use of the chemical potential is a way to impose a given expectation value for some conserved
charge. The alternative option of selecting a given sector of the charge in the path integral presents
additional difficulties, but something can be learned by a series expansion of the fermion
determinant~\cite{Palu3}. To be concrete we will refer to  the case of QCD, but the method  can be applied
to other cases with appropriate modifications.

In the absence of any condition on the baryon number the quark determinant is
\be
\det \, Q = \int [d \overline{q}\, d q] \exp S_q,
\ee
where $S_q$ is the quark action and $Q$ the quark matrix. My strategy is to write $ \det \, Q $ as the
trace of the transfer matrix acting in the quark Fock space, impose the restriction to a given baryonic
sector,  and then rewrite the trace as the determinant of a modified quark matrix. The round trip is done
by  mapping the Grassmann algebra generated by the quark fields into the Fock space  following the
construction of L\"uscher ~\cite{Lusc}. But while his paper is based on the mere existence of the map, 
in order to enforce the projection I will make use of a concrete realization by means of coherent states.

 The first step is then to write the unconstrained determinant as a trace in the Fock space
\be
\det \, Q= \mbox{Tr} \, \hat{\cal{T}}.  \label{trace} 
\ee
The second step is to impose the restriction to a sector with baryon number $n_B$ by inserting in the 
trace the appropriate  projection operator $\hat{P}_{n_B} $
\be
 \det \, Q|_{n_B} = \mbox{Tr} \left( \hat{\cal{T}} \hat{P}_{n_B} \right) \int [dx^+ dx\, dy^+ dy ]
 \exp \left(- x^+x-y^+y \right)
\<x,y|\hat{\cal{T}} \hat{P}_{n_B}|-x,-y\>  \label{Berezin}
\ee
which will be expressed in terms of the determinant of a modified quark matrix. 
The kernel $\<x,y|\hat{\cal{T}} \hat{P}_{n_B}|-x,-y\> $ has the integral form
\be
 \<x,y|\hat{\cal{T}}\hat{ P}_{n_B}|-x,-y\> = \int [dz^+ dz\, dw^+ dw]
\exp \left(- z^+z-w^+w \right) 
 \<x,y|\hat{\cal{T}}|z,w\>
 \<z,w| \hat{P}_{n_B}|-x,-y\> . \label{convol}
\ee

The expression of the kernel $\<x,y|\hat{\cal{T}}|z,w\>$ will not be reported here, while 
that of $\hat{P}_{n_B}$ can easily be derived 
\be
 \< z,w|\hat{P}_{n_B}|-x,-y\> = \sum_{r=0}^{\infty}(-1)^{n_B}
 \<(\hat{y} w^+)^r (\hat{x} z^+ )^{(n_B+r)} 
(x \hat{x}^+ )^{n_B+r} (y \hat{y}^+)^r \>
 {1 \over ((n_B +r)! r!)^2}.
\ee
 Since ${\hat x}^+ , {\hat y}^+$ are creation operators of quarks and antiquarks
respectively, we see that
the r-th term of this  series gives the gauge-invariant contribution of $n_B$ valence 
quarks plus $r$ quark-antiquark pairs. 
 
Needless to say, for $n_B=0$, $ \det \, Q|_{n_B}$ does not reduce to the
unconstrained determinant: Indeed also baryonic states are present in the
unconstrained
determinant, while they are absent in $ \det \, Q|_{n_B=0}$. In QCD at nonvanishing
temperature it makes a
difference  whether we impose or we do not the condition $n_B=0$. 
In view of the
relatively low value of the critical temperature with respect to the nucleon mass, however,
we do not expect significant effects from the restriction to a given baryon sector unless we go to
exceedingly
 high temperatures.

Let me now proceed to derive our final result. By evaluating
the vacuum expectation values appearing in the last equation we express the kernel 
of the projection  operator in terms of Grassmann variables only
\be
  \< z,w|\hat{P}_{n_B}|-x,-y \>  = \sum_{r=0}^{\infty}(-1)^{n_B}
{1 \over (n_B +r)! r!} 
(z^+ x)^{n_B+r} (w^+ y)^r.
\ee
To evaluate the integral of Eq.\reff{convol} I rewrite the above
equation in exponential form
\be
 \< z,w|\hat{P}_{n_B}|x,y \> = \sum_{r=0}^{\infty}{1 \over (n_B +r)! r!}
{\partial^{n_B+r} \over  \partial j_1^{n_B+r} }
 {\partial^r \over \partial j_2^r } 
\exp(-j_1 z^+x - j_2 w^+ y ) |_{j_1=j_2=0}.
\ee
The integrals of Eqs.\reff{Berezin},\reff{convol} are Gaussian and we get the
constrained determinant in terms of the determinant of a modified quark matrix
\be
 \det \, Q|_{n_B} =  \sum_{r=0}^{\infty}{1 \over (n_B +r)! r!} {\partial^{n_B+r} \over 
\partial j_1^{n_B+r} } {\partial^r \over \partial j_2^r }
\det\left( Q + \delta Q_1 + \delta Q_2 \right) |_{j_1=j_2=0}.
\ee
The explicit form of the variations $ \delta Q_1 , \delta Q_2 $ of the quark matrix is
not important here, but we warn the reader that there is an error in their expression in ref.\cite{Palu2}.

\section{Bosonization in many-body systems and relativistic field theories }

The low energy collective excitations of many-fermion systems can be described by effective bosons. Well
known examples are the Cooper pairs of Superconductivity, the bosons of the Interacting Boson
Model of Nuclear Physics, the chiral mesons and the quark pairs of color superconductivity in QCD. In all
these cases the effective  bosons are generated by attractive interactions, but effective bosons can arise
also in the  presence of repulsive forces, like in the Hubbard model~\cite{Cini}. Some of the effective
bosons are Goldstone  bosons, and then there is a general theory which tells that they live in the coset
space of the  group which is spontaneously broken and dictates how they are related to the original
fields~\cite{Wein}.
 But there is no  general procedure to reformulate the fermionic theory in terms of the effective bosonic
degrees  of freedom, even though there are several recipes for specific cases which are reviewed
in~\cite{Klein}. A more flexible approach is based on the Hubbard--Stratonovich transformation which 
linearizes the fermionic interaction by introducing bosonic auxiliary fields which are then promoted  to
physical life. The typical resulting structure is that of chiral theories~\cite{Mira}. But in such an
approach an energy scale emerges naturally, and only excitations  of lower energy can be described by the
auxiliary fields. Moreover in renormalizable relativistic  field theories like QCD, the fermion Lagrangian
is quadratic to start with, so that the Hubbard--Stratonovich transformation cannot be used. One can
add quartic interactions as irrelevant operators, and this can help in numerical simulations, but
has not led so far to a formulation of low energy QCD in terms of chiral mesons.

I present a new approach~\cite{Palu3} to bosonization which does not suffer from the above limitations and
can be applied to theories with quartic and quadratic interactions as well. It is based on the 
evaluation of the partition function restricted to the bosonic composites of interest. By rewriting 
the partition function so obtained in functional form we get the euclidean action of the composite 
bosons from which in the nonrelativistic theories we can derive the Hamiltonian. In this way I
derived the Interacting Boson Model from a Nuclear Hamiltonian. In the case  of pure pairing, I reproduce
the well known  results for the excitations corresponding to the addition and removal of pairs of fermions,
as well as for the seniority excitations which are inaccessible by the Hubbard--Stratonovich
method. Indeed at least in this example this theory does not have the structure of a chiral
expansion. 

For the relativistic case an investigation is in progress~\cite{Cara}.

Let me start by defining the composites in terms of the fermion operators ${\hat c}$
\be
{\hat b}^{\dagger}_J= { 1\over 2} {\hat c}^{\dagger} B^{\dagger}_J {\hat c}^{\dagger}
= { 1\over 2} \sum_{m_1,m_2}{\hat c}^{\dagger}_{m_1}\left( B^{\dagger}_J\right)_{m_1,m_2}
 {\hat c}^{\dagger}_{m_2}.
\ee
In the above equation $m$ represents all the fermion intrinsic quantum numbers and position 
coordinates and $J$ the quantum numbers of the composites. I assume all the structure
matrices $B_J$ to have one and the same dimension which I denote by $2 \, \Omega$. The fermionic
operators have canonical anticommutation relations, while for the composites
\be
[ {\hat b}_{J_1},{\hat b}_{J_2}^{\dagger}] = { 1\over 2}\mbox{tr} (B_{J_1} B^{\dagger}_{J_2} ) - 
{\hat c}^{\dagger} B^{\dagger}_{J_2} B_{J_1} {\hat c}.
\ee
It is then natural to require  the normalization
\be
\mbox{tr} (B^{\dagger}_{J_1}, B_{J_2} ) = 2 \delta_{J_1,J_2}.
\ee

A convenient way to get the euclidean path integral from the trace of the transfer
matrix is to use coherent states of composites. Therefore I introduce the  operator
\be 
{\cal P} = \int db^* db (\< b| b \>)^{-1}|b \> \< b|
\ee
where
\be
|b \> = |\exp (b \,\cdot {\hat b}^{\dagger})\>.
\ee
I adopted the convention
\be
b \,\cdot {\hat b}^{\dagger} = \sum_J b_J {\hat b}^{\dagger}_J.
\ee
If the ${\hat b}$'s where operators of elementary bosons ${\cal P}$ would be the identity in the boson Fock
space. I would like ${\cal P} $ to be the identity in the fermion subspace of the composites. To see
the action of ${\cal P} $ on  composite operators let us first consider the case where there is only one
composite with structure function satisfying the equation
\be
B^{\dagger} B= { 1\over \Omega}1\!\!1. \label{neces}
\ee
Then we find
\be
\<b|b\>^{-1}= \left(1+ {1\over \Omega} b_1^* b \right) ^{-\Omega}
\ee
and
\be
\<b_1|( {\hat b}^{\dagger})^n \> = C_n (b^*)^n,
\ee
where
\be
C_n  = { \Omega! \over (\Omega-n)! \Omega^n}= 
\left( 1- { 1\over \Omega}\right) \left( 1- { 2\over \Omega}\right)...
 \left( 1- { n-1\over  \Omega}\right). 
\ee
Now we can determine the action of ${\cal P}$ on the composites
\be
{\cal P}|({\hat b}^{\dagger})^n \> =  \left( 1- { n \over \Omega}\right)^{-1} 
\left( 1 - { n+1 \over \Omega}\right)^{-1} |({\hat b}^{\dagger})^n \>
\ee
which shows that ${\cal P}$ behaves approximately like the identity with an error of the order of 
$n/ \Omega$. It is perhaps worth while noticing that in
the limit of infinite $\Omega$ we recover exactly the expressions valid for elementary bosons, in
particular
\be
\<b_1|b\> =  \left(1+ {1\over \Omega} b_1^* b \right) ^{\Omega} \rightarrow  \exp ( b_1^* b),
\,\,\, \Omega \rightarrow \infty.
\ee

It might appear that the treatment of states with $n \sim \Omega$ is precluded, but this is not true.
 Indeed if we are
interested in states with $n = {\overline n} + \nu$ for an arbitrary reference state ${\overline n}$,
  we redefine ${\cal P}$ according to
\be
{\cal P}_{{\overline n}} = { ( \Omega - {\overline n})^2 \over \Omega^2} {\cal P}_0.
\ee
We then have
\be
 {\cal P}_{{\overline n}}|({\hat b}^{\dagger})^n \>= 
\left( 1 - { \nu \over \Omega - {\overline n}} \right)^{-1}
 \left( 1 - { \nu +1 \over \Omega - {\overline n}} \right)^{-1}  |({\hat b}^{\dagger})^n \>
\ee
which shows that ${\cal P}_{{\overline n}} $ behaves like the identity in the neighborhood
of the reference state up to an error of order $ \nu /(\Omega - {\overline n})$, namely the measure 
$\<b|b\>^{-1}$ is essentially uniform. 

In the general case of many composites we have
\be
\< b_1|b \> = \left[ \det \left( 1\!\!1 + \beta_1 \beta \right) \right]^{1 \over 2}
\ee
where the matrix $\beta$ is
\be
\beta = b \cdot B^{\dagger}
\ee
and
\be
\<b_1| ({\hat b_{I_0}}^{\dagger})^{n_0}...({\hat b_{I_i}}^{\dagger})^{n_i} \> =
{\partial^{n_0} \over \partial x_0^{n_0} }... {\partial^{n_i} \over \partial x_i^{n_i} }
 \exp \{{ 1\over 2} \mbox{Tr} \ln [ 1\!\!1 + (x \cdot B^{\dagger})( b^*_1 \cdot B)] \}|_{x=0}.
\ee
We must now make an assumption which replaces Eq.~\ref{neces}, namely that all the eigenvalues of the
matrices
$ B_J^{\dagger}B_J$ are much smaller than $\Omega$. Then we find again that ${\cal P}$ approximates
the identity with an error of order $ 1 / \Omega$.

Now we are equipped to carry out the program outlined at the beginning. The 
first step is the evaluation of the partition function $Z_C$ restricted to fermionic composites. To
this end we divide the inverse temperature in $N_0$ intervals of spacing $\tau$
\be
{ 1 \over T}  =  N_0 \tau
\ee
and write
\be
Z_c = \mbox{tr} \left( {\cal P} {\cal T} \right)^{N_0}
\ee
where ${\cal T}$ is the transfer matrix. In the nonrelativistic case ${\cal T}$ is expressed in terms of the
Hamiltonian
\be
{\cal T}= \exp \left(- \tau {\hat  H}\right),\,\,\,
\ee
and the Hamiltonian is the generator of continuous time translations. In relativistic  field
theories instead only the transfer matrix is known in general, and the above equation can be used to 
define a Hamiltonian, but only as the generator of discrete translations by the time spacing $\tau$.

 At this point  we must evaluate the matrix element $\<b_1| {\cal T} |b\>$ and to do this we must distinguish
between relativistic field theories and many-body systems. In the first case the transfer matrix is a product of
exponentials of quadratic forms in the fermion operators~\cite{Lusc}, and the matrix element can be directly and
exactly evaluated~\cite{Cara}.  In
the second  case
one must at an intermediate stage expand with respect to the time spacing $\tau$. This does not
introduce any error because one can retain all the terms which give finite contributions
in the limit $\tau \rightarrow 0$.
We will report here only the nonrelativistic calculation. The most general Hamiltonian can be written
\be
{\hat H}= {\hat c}^{\dagger} h_0 \, {\hat c} -
\sum_K g_K { 1\over 2} \, {\hat c}^{\dagger} F_K^{\dagger} {\hat c}^{\dagger}
{ 1\over 2}\, {\hat c}\, F_K \, {\hat c} \, ,
\ee
where $K$ represents all the necessary quantum numbers. The single-particle term includes the 
single-particle energy with matrix $e$, any single-particle interaction with external fields described by
the matrix ${\cal M}$ and the chemical potential $\mu$
\be
h_0= e + {\cal M} - \mu.
\ee
Therefore we will be able to solve the  problem of fermion-boson mapping by determining the
interaction of the composite bosons with external fields. Assuming for the potential form factors the
normalization
\be
\mbox{tr} ( F_K^{\dagger} F_K) = 2 \, \Omega \
\ee
and setting
\be
\Gamma_t= \left( 1\!\!1 + \beta_t^* \beta_{t-1}\right)^{-1}
\ee
 we get the  euclidean action 
\begin{eqnarray}
 S  &=& \tau \sum_t \left\{ { 1\over 2 \tau} \mbox{tr} [ \ln (1\!\!1 + \beta_t^* \beta_t) 
- \ln \Gamma_t] - H_1   + { 1\over 4}  \sum_K g_K \mbox{tr} \left[   \mbox{tr} (\Gamma_t \beta_t^* 
F_K^{\dagger}) \,
\mbox{tr}(\Gamma_t F_K \beta_{t-1}) \right. \right.
\nonumber\\
& & \left. \left. -  2 \, \mbox{tr} \left( \Gamma_t  F^{\dagger}_K F_K \right)
- \mbox{tr} [ \Gamma_t \beta_t^* F_K^{\dagger}, \Gamma_t  F_K \beta_{t-1}]_+   
  \right]   + { 1\over 2}  \mbox{tr} \left[ \beta_t^*  
\, ( \beta_{t-1} \, h^T + h \,\, \beta_{t-1} ) \right]   \right\} ,
\end{eqnarray}
where $[..,..]_+$ is an anticommutator.
This action differs from that of elementary bosons because

i) the time derivative terms (contained in the first line) are non canonical

ii) the coupling of the chemical potential (which appears in $h$) is also noncanonical, since
it is not quadratic in the boson fields

iii) the function $\Gamma$ becomes singular when the number of bosons is of order $\Omega$, which reflects the
Pauli principle.
 
We remind the reader that the only approximation done concerns the operator ${\cal P}$. Therefore
these are to be regarded as true features of compositeness.

The bosonization of the system we considered has thus been accomplished. In particular the
fermionic interactions with external fields can  be expressed in terms of the bosonic terms
which involve the matrix ${\cal M}$ (appearing in $h$) and the dynamical problem of the interacting
(composite) bosons can be solved within the path integral formalism. Part of this problem is the 
determination of the structure matrices $B_J$. This can be done by expressing the energies in terms of
them and applying a variational principle which gives rise to an eigenvalue equation.

The Hamiltonian can be derived by standard procedures.

\section{Conclusion}

I showed that there is a number of problems which can be easily dealt with in the operator
form of the partition function. Only afterwards it can be given the functional form of the path
integral which is more convenient for many purposes. 

There are other examples I left over for different reasons. Among these I would like to mention the problem
of the restriction of gauge theories to physical states, which has not yet found a general and
satisfactory solution.


\end{document}